\documentclass[pre,showpacs,amssymb]{revtex4}

\usepackage{amsmath,graphicx}
\usepackage{epsfig}

\begin{document}

\title{Fractional Brownian motion approach to polymer translocation:\\ 
the governing equation of motion}
\author{J. L. A. Dubbeldam$^{2}$, V. G. Rostiashvili$^1$, A. Milchev$^{1,3}$,
and T. A. Vilgis$^1$}
\affiliation{$^1$ Max Planck Institute for Polymer Research
  10 Ackermannweg 55128 Mainz, Germany.\\
$^2$ Delft University of Technology 2628CD Delft, The Netherlands\\
$^3$ Institute for Physical Chemistry Bulgarian Academy of Sciences, 1113 Sofia,
Bulgaria
}

\begin{abstract}
We suggest a governing equation which describes the process of polymer chain
translocation through a narrow pore and reconciles the seemingly contradictory
features of such dynamics: (i) a Gaussian probability distribution of the
translocated number of polymer segments at time $t$ after the process has begun,
and (ii) a sub-diffusive increase of the distribution variance $\Delta (t)$ with
elapsed time, $\Delta(t) \propto t^{\alpha}$.  The latter quantity measures the
mean-squared number $s$ of polymer segments which have passed through the pore,
$\Delta(t) = \langle [s(t)-s(t=0)]^2\rangle$,  and is known to grow  with an
anomalous diffusion exponent $\alpha < 1$.

Our main assumption - a Gaussian distribution of the translocation velocity
$v(t)$ - and some important theoretical results, derived recently, are shown to
be supported by extensive Brownian dynamics simulation which we performed in
$3D$. We also numerically confirm the predictions made in  ref.\cite{Kantor_3},
that the exponent $\alpha$ changes from $0.91$  to $0.55$, to $0.91$, for short,
intermediate and long time regimes, respectively.
\end{abstract}


\maketitle

\section{Introduction}

Polymer translocation has attracted recently a lot of attention as it plays a
crucial role in processes such as DNA and RNA transport through membrane
channels \cite{Meller} . Ultimately, this research might lead to a longstanding
objective of nucleotic transport: fast and cheap sequencing of DNA. Meanwhile
many interesting experimental and theoretical findings have been reported
\cite{Zwolak}. A lot of important observations have been gained by means of
computer simulations
\cite{Luo,Huopaniemi,Dubbeldam_1,Dubbeldam_2,Panja_1,Panja_2,Panja_3}. However,
notwithstanding the significance of translocation phenomena both as possible
technological application and from the standpoint of basic research, the
understanding of the polymer translocation through a narrow pore is still
elusive and in some respects controversial \cite{Sakaue}.

In most theoretical descriptions (except Refs. \cite{Sakaue}, where a full
Cartesian-space treatment has been suggested) the translocation process is
assumed to be captured by a single variable, the translocation coordinate
$s(t)$, which measures the number of translocated segments at time $t$.
Generally, $s(t)$ would depend on the external potential, if present, and be
influenced by random noise. At present it is well established that the
translocation coordinate $s(t)$ follows an anomalous, sub-diffusive law,
although the physical background as well as the equation which governs the
probability distribution function (PDF) $W(s, t)$ that $s$ segments have passed
through the pore at time $t$ are still controversial.

It was suggested recently \cite{Metzler} that the translocation dynamics is
governed by a {\it fractional} diffusion equation (FDE) with an (anomalous)
diffusion exponent $\alpha < 1$. Thus a description of the translocation process
in terms of $W(s,t)$ was derived \cite{Dubbeldam_1} which appeared to agree
favorably with Monte
Carlo simulation results. This approach was generalized \cite{Dubbeldam_2} to
the case of biased translocation, driven by external potential, which was
described by an appropriate fractional Fokker-Planck equation (FFPE). As for the
non-driven translocation, however, the FFPE approach yields a {\em non-Gaussian}
distribution $W(s,t)$ whose width $\Delta (t)$, diverges sub-diffusionally with
the elapsed time, $\Delta (t) \propto t^\alpha$. While this latter property may
readily be verified in numeric experiments, the unambiguous determination of the
precise functional shape of $W(s,t)$ is rather difficult due to progressively
deteriorating statistics of the distribution at late times.

Meanwhile, several recent publications\cite{Kantor_1,Kantor_2,Kantor_3}, devoted
to the unbiased translocation dynamics of a Gaussian $1D$ chain \cite{Kantor_1}
as well as to that of a $2D$ self-avoiding chain \cite{Kantor_2}, have validated
the sub-diffusive behavior of $\langle s^2(t)\rangle$ with an exponent $\alpha
\simeq 0.8$. Nonetheless, these new findings cast serious doubts as to whether
the FFPE provides indeed an adequate description of nondriven translocation
dynamics:
\begin{itemize}
 \item it was found \cite{Kantor_1,Kantor_2,Kantor_3} by means of computer
simulations that the probability distribution $W(s, t)$ of the translocation
coordinate $s$, {\em stays Gaussian} for different time moments not exceeding
the mean translocation time $\langle \tau \rangle$. At larger times this
distribution attains a more complex form.

 \item  the long time tail of the first-passage-time (FPT) distribution was
found to be {\em exponentially decreasing} thus challenging the power-law
behavior, suggested earlier \cite{Dubbeldam_1} within the framework of the FFPE.

\end{itemize}
These findings question the validity of the FDE approach in
the description of translocation dynamics. Indeed, one cannot reconcile the
aforementioned controversial features of the translocation dynamics within a
FFPE description.

In the present work we revisit the problem and demonstrate that a proper
Fokker-Planck equation of motion (FPEM) which governs the PDF $W(s, t)$ and
faithfully reproduces all recently found properties of translocation dynamics
may be rationalized and solved within the framework of fractional Brownian
motion (fBm) \cite{West_Book}. Namely, we treat the definition of the
translocation coordinate velocity, $v(t) = d s(t)/d t$, as a Langevin equation,
where the velocity auto-correlation function, $G(t) = \langle v(t) v(0)\rangle
$, is considered as dynamic input without making any {\em a priori} assumption
about the dynamics that underlies the translocation process. In principle,
$G(t)$ may be taken from the computer simulation. Such an approach finally leads
to a FPEM (with a time-dependent diffusion coefficient) for $W(s, t)$ in
unbiased and biased cases. We make an extensive Brownian dynamics simulation
study (using GROMACS simulation package) in order to check and justify our
analytical results.

In Section \ref{chain_dynamics} we derive a governing equation for PDF
$W(s,t)$, calculate its first moments, and present an exact analytic solution
for the first passage time distribution of translocation times as well as for
the so called {\em survival} probability in terms of monomer consecutive number
$s$ and elapsed time $t$. A comparison with our Brownian Dynamics simulation
data is performed in Section   \ref{sim}. Then our main conclusions are given as
a brief summary in Section \ref{summary}.

\section{Chain Translocation Dynamics}
\label{chain_dynamics}

\subsection{Fokker-Planck equation with a time-dependent drift and diffusion coefficient}

\begin{figure}[htb]
\includegraphics[scale=0.5]{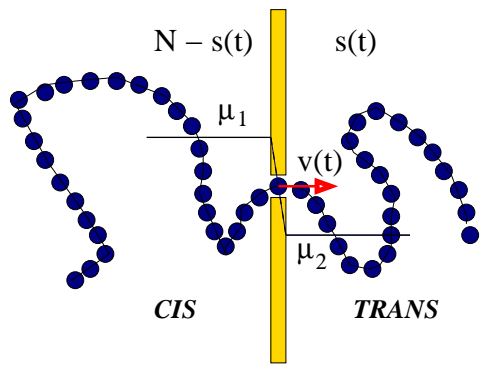}
\caption{(Color online) Chain translocation through a nanopore with an instantaneous
translocation coordinate $s(t)$. The translocation velocity is approximated by
the velocity of the bead which resides inside the pore. The separating membrane
is taken as sufficiently thin so that no more than one bead may be in the pore
at a time. The driving force caused by the chemical potential gradient within
the pore, $f = (\mu_1 - \mu_2)/T$, acts on the same bead. $\mu_1$ and $\mu_2$
denote values of the chemical potential on both sides of the separating
membrane.}
\label{Fold}
\end{figure}

One may derive the FPEM for the distribution $W(s, t)$, starting from the
Langevin equation,
\begin{equation}
\frac{d}{d t} s(t) = v(t),
\label{Langevin}
\end{equation}
where, by assumption, the translocation velocity $v(t)= ds(t)/dt$ follows
Gaussian statistics (i.e., is Maxwell-Boltzmann distributed). If a (generally
time dependent) external driving force $f(t)$ is present, the mean velocity
$\langle v(t)\rangle = f(t)/\xi_0$ where $\xi_0$ denotes the friction
coefficient. Note that the velocity $v(t)$ in eq.~(\ref{Langevin}) reflects the
change in the $s$ coordinate per unit time. By measuring $s(t)$ as the contour
length of the chain on the {\em trans}-side of the membrane, a reasonable
estimate for the translocation velocity is obtained. In fact, $v(t)$ can be
approximated by the velocity of the bead inside the pore in the direction perpendicular
to the wall, which we denote by $v_z(t)$ (see Fig.~\ref{Fold}). Thus we tacitly
assume that translocation velocity may be faithfully characterized by bead
velocity in Cartesian space. The correctness of this approximation will be
addressed in Section \ref{sim}.

The variance of $s(t)$ can easily be calculated under the assumption that the
velocity auto-correlation function $\langle v(t_1) v(t_2)\rangle$, defined as
$\langle v(t_1) v(t_2)\rangle = \lim_{T\rightarrow \infty} (1/T) \int_{0}^{T}
v(t_{1}+t) v(t_{2}+t) dt$ depends on the difference $|t_{1} - t_{2}|$ only
(assuming time translational invariance). This leads to the relationship
\cite{Boon}
\begin{equation}
\left\langle [s - s_{0}]^2 (t)\right\rangle = 2 \int_{0}^{t} (t - t')
\left\langle v(t') v(0)\right\rangle d t'. \label{Velocity}
\end{equation}
where $s_0 \equiv s(t=0)$ is the initial value of the translocation coordinate.
In the case of biased translocation, the driving force is usually due to a
chemical potential gradient $\Delta \mu$, which is typically generated by
applying a voltage difference across the membrane, i.e., $f = \Delta \: \mu/T$,
where $\Delta \: \mu = \mu_1 - \mu_2$. Here $f$ is imposed on the monomer which
is currently in the pore (cf. Fig. \ref{Fold}). As pointed out above, the
velocity $v(t)$ is assumed to be a Gaussian random variable with the first two
cumulants given as
\begin{equation}
 \langle v(t)\rangle = \dfrac{f(t)}{\xi_0},
\label{First}
\end{equation}
and
\begin{equation}
 G(t_1, t_2) \equiv \langle [v(t_1) - \langle v(t_1)\rangle][v(t_2) - \langle v
(t_2)\rangle]\rangle. \label{Second}
\end{equation}

We next consider the PDF $W(s, t)$, which is defined as
\begin{equation}
W(s, t) =  \Bigl<\delta (s - s_0 - \int\limits_{0}^{t} v(t') dt') \Bigr>,
\label{W}
\end{equation}
and satisfies the initial condition $W(s,0)=\delta(s-s_0)$. It is
straightforward to show that
\begin{eqnarray}
\frac{\partial}{\partial t}  \: W(s, t) &=&- \frac{\partial}{\partial s} \:
\Bigl[\left\langle v(t) \right\rangle \: \Bigl< \delta (s - s_0 -
\int\limits_{0}^{t} v(t') dt')\Bigr> \Bigr]-
\frac{\partial}{\partial s} \: \int\limits_{0}^{t} \: d\tau G(t,\tau) \: \Bigl<
\frac{\delta}{\delta v(\tau)} \: \delta (s - s_0 - \int\limits_{0}^{t} v(t')
dt')\Bigr>,
\label{PDF}
\end{eqnarray}
where we have  used Novikov's theorem \cite{Novikov,Zinn} with the two first
cumulants of $v(t)$ given by eqs. (\ref{First}) and (\ref{Second}). In the last
term in eq. (\ref{PDF}) we have also used $\langle(\delta/\delta v(\tau)) \:
\delta (s - s_0 - \int\limits_{0}^{t} v(t') d t') \rangle = - (\partial/\partial
s) \langle \delta (s - s_0 -\int\limits_{0}^{t} v(t') d t')\rangle$ as well as
$\int_{0}^{t} \delta (\tau - t')dt'=1$ for $\tau \le t$. Eventually the FPEM
takes on the form:
\begin{eqnarray}
 \frac{\partial}{\partial t}  \: W(s, t) =  - \frac{f (t)}{\xi_0} \:
\frac{\partial}{\partial s} \:W(s, t)+  {\cal D}(t) \frac{\partial^2}{\partial
s^2}  \: W (s, t).
\label{FP}
\end{eqnarray}
with a time-dependent diffusion coefficient ${\cal D}(t) = \int_{0}^{t}
G(t,\tau) d \tau$. This equation of motion is different from the FFPE
\cite{Metzler1} which was used in our previous investigation
\cite{Dubbeldam_1,Dubbeldam_2}. In contrast to FFPE, it is Markovian but
contains a time-dependent drift term and a time-dependent diffusion coefficient.
Moreover, the form of eq .(\ref{FP}) does not depend on the underlying dynamics
(underdamped  or overdamped regimes!) as far as the driving force $f(t)$ and the
velocity correlator $G(t_1, t_2)$ are not specified. We should emphasize that
Eq.~(\ref{FP}) has been by assuming that the effective friction coefficient
$\xi_0$ remains constant with time. This is supported by computer simulation
data, cf. Fig. 6a in Bhattachrya et al.~\cite{Aniket}. I was shown~\cite{Aniket}
that for a driven translocation the mean center-of-mass velocity quickly
saturates to a constant value after a short initial increase. The transient
regime itself lasts less than 10\% of the total translocation time for a short
chain ($N = 128$) and quickly vanishes for longer chains. Recently the
translocation dynamics has been treated \cite{Cherayil_1,Cherayil_2} on the
basis of a FPEM with a time dependent drift (even for a {\em constant} driving
force) and diffusion coefficient.

In the simplest case of an unbiased process (i.e., $f=0$), the system is in a
steady state and ${\cal D}(t) = \int_{0}^{t} G(t-\tau) d \tau = \int_{0}^{t}
G(\tau) d \tau$. Then with the boundary conditions at infinity eq.(\ref{FP}) may
be solved by defining a new time variable $\theta$
\begin{eqnarray}
 \frac{d \theta}{d t} =   \int\limits_{0}^{t} d t' \: G(t'),
\label{Theta}
\end{eqnarray}
i.e., $\theta(t) =  \int\limits_{0}^{t} d t' \: \int\limits_{0}^{t'} d
t'' \: G(t'') =  \int\limits_{0}^{t} (t - t') G(t')  d t'$. A comparison
with eq.(\ref{Velocity}) then yields
\begin{eqnarray}
 \theta (t) = \dfrac{1}{2} \left\langle(s - s_0)^2\right\rangle \propto t^{\alpha},
\label{Comparison}
\end{eqnarray}
where we have assumed that the sub-diffusive behavior of $\langle [s - s_0]^2
(t)\rangle$ is described by an exponent $\alpha < 1$. Thus eq. (\ref{FP}) takes
on the form
\begin{eqnarray}
  \frac{\partial}{\partial \theta}  \: {\tilde W}(s, \theta) =
\frac{\partial^2}{\partial s^2} \: {\tilde W}(s, \theta).
\label{Markov_Equation}
\end{eqnarray}
The solution of eq.~(\ref{Markov_Equation}) reads ${\tilde W}(s, \theta)= \left(
1/2\pi \theta\right)^{1/2} \exp\left[ - (s-s_0)^2/4\theta\right] $. Then, after
taking into account eq.(\ref{Comparison}), one finds
\begin{eqnarray}
W(s, t) = \frac{1}{\sqrt{2\pi D t^{\alpha}}} \: \exp\left[ -
\frac{(s - s_0)^2}{4 Dt^{\alpha}}\right],
\label{Final_Solution}
\end{eqnarray}
where $D$ is a constant. This result for $W(s,t)$ has been obtained very
recently by Panja \cite{Panja_arXiv,Panja_JSMP} in a more general context
dedicated to anomalous polymer dynamics.

Evidently, the distribution $W(s, t)$ is indeed Gaussian albeit with a width
proportional to the second moment that grows sub-diffusively with elapsed time.
This result reproduces the Monte-Carlo simulation findings, reported recently by
Kantor and Kardar \cite{Kantor_1,Kantor_2}. It should be noted that such a
process, described by a Gaussian distribution with anomalous width is sometimes
referred to as {\em fractional Brownian motion} \cite{West_Book,West_1} and has
been discussed in the context of DNA sequence statistics \cite{West_2}.

\subsection{Boundary conditions}

Eq.~(\ref{Markov_Equation}) should be considered along with the appropriate
boundary conditions. In the special case when the translocation starts from the
middle monomer (i.e., $s_0 = N/2$) and stops as soon as any of the two chain
ends passes the pore, both boundary conditions are adsorbing, i.e., ${\tilde
W}(s=0, \theta) = 0, \; {\tilde W}(s = N, \theta) = 0$. The corresponding
initial condition reads ${\tilde W}(s, \theta = 0) = \delta (s - s_0)$. Recall
that $\theta (t=0) = 0$.

The full solution can be represented as a sum over eigenfunctions $\varphi_n
(s)$, i.e., ${\tilde W}(s, \theta ) = \sum_{n=0}^{\infty} T_n(\theta) \varphi_n
(s)$ (see, e.g., \cite{Risken}) where $\varphi_n (s)$ obey the equations $(d^2/d
s^2) \varphi_n (s ) + \lambda_n \varphi_n (s) = 0$. The temporal part is
governed by the equation $(d / d \theta ) T_n(\theta) + \lambda_n T_n(\theta) =
0$ with eigenvalues $\lambda_n$. The proper eigenfunctions which satisfy both
boundary conditions, are $\varphi_n (s) = A_n \sin (n \pi/N)$ (the factor $A_n$
can be fixed by the initial condition). The eigenvalues are then $\lambda_n =
(\pi n/N)^2$. The completeness relation for the eigenfunctions (see Sec. 5.4 in
\cite{Risken} ) has in this case the form $ \delta (s - s_0) = \dfrac{2}{N} \:
\sum\limits_{n=0}^{\infty} \: \sin \left(\dfrac{n \pi s}{N} \right)
\sin \left( \dfrac{n \pi s_0}{N} \right)$ so that the full solution becomes thus
\begin{eqnarray}\label{sol_eq}
 {\tilde W}(s, \theta ) = \dfrac{2}{N} \: \sum\limits_{n=0}^{\infty} \: \sin
\left(\dfrac{n \pi s}{N} \right) \sin\left(\dfrac{n \pi s_0}{N} \right) \: \exp
\left[ - \left(\dfrac{n \pi}{N} \right)^2 \theta \right].
\end{eqnarray}
Note that eq. (\ref{Comparison}) defines a sub-diffusional law $\theta (t) = D
\: t^{\alpha}$,  which  governs the solution, eq. (\ref{sol_eq}).
Eventually, if the starting point $s_0 = N/2$ (as is the case in our
MD-simulation), only the odd terms in the series, eq. (\ref{sol_eq}), survive,
i.e., $n = 2 m + 1$ and the final result reads:
\begin{eqnarray}
{\tilde W} (s, t ) = \dfrac{2}{N} \: \sum\limits_{m =0}^{\infty} \:
(- 1)^m \: \sin\left[\dfrac{(2 m + 1) \pi s}{N} \right] \: \exp
\left[ - \dfrac{(2 m + 1)^2 \pi^2}{N^2}   D \: t^{\alpha} \right].
\label{Solution_Final}
\end{eqnarray}

The first two moments generated by this probability distribution contain a lot
of information, which  permits us to  compare  our theoretical results with
those from literature and with the numerical results of Section \ref{sim}.

\subsection{First and second moments of $W(s, t )$}

Taking into account that $\int_{0}^{N} \sin [(2 m + 1) \pi s/N] d s = 2
N/(2 m + 1)\pi$ and $\int_{0}^{N} s \sin [(2 m + 1) \pi s/N] d s =
N^2/(2 m + 1)\pi$,  one obtains for the first moment $\langle s \rangle =
 \int\limits_{0}^{N} \: s W(s,t)d s/\int\limits_{0}^{N} \: W (s, t )  d s = s_0
= N/2$. The second moment  is centered at $s_0$, hence, $\Delta \equiv \langle
s^2 \rangle - \langle s \rangle^2 = \int\limits_{0}^{N} \: [s-s_0]^2  W (s, t )
d s / \int\limits_{0}^{N} \: W (s, t )  d s $. This leads to the following
expression for the variance,
\begin{eqnarray}
 \Delta \equiv \langle  s^2 \rangle - \langle s \rangle^2 = \dfrac{N^2}{4}
\left\{ 1 -
\dfrac{8\sum\limits_{m=0}^{\infty} \: \dfrac{(- 1)^m}{(2 m + 1)^3} \exp\left[ -
\dfrac{(2 m + 1)^2 \pi^2}{N^2} D \:  t^{\alpha} \right]}{\pi^2
\sum\limits_{m=0}^{\infty}\dfrac{(- 1)^m}{(2 m + 1)}  \exp\left[ - \dfrac{(2 m +
1)^2 \pi^2}{N^2} D \:  t^{\alpha} \right]} \right\}.
\label{Second_Moment}
\end{eqnarray}
At $t \to \infty$  the second moment $\Delta$ reaches a plateau, i.e.,
$\Delta =
(N^2/4) (1 - 8/\pi^2)$. It can be readily shown (taking into account that
$\sum_{m=0}^{\infty} (-1)^m/(2m+1)= \pi/4$ and $\sum_{m=0}^{\infty}
(-1)^m/(2m+1)^3= \pi^3/32$) that $\Delta (t=0) = 0$. The time dependence of the
second moment $\Delta$, given by eq.~(\ref{Second_Moment}), will be discussed
further in the Section \ref{sim}.

\subsection{First passage time distribution}

The first passage time distribution (FPTD) function is defined as \cite{Risken}
and describes the probability distribution of the observed translocation times:
\begin{eqnarray}
 Q (t) = - \dfrac{d}{d t} \: \int\limits_{0}^{N} \: W (s, t) d s.
\label{FPT_Definition}
\end{eqnarray}
Taking into account eq. (\ref{Solution_Final}), this definition yields
\begin{eqnarray}
 Q (t) = \dfrac{4\pi \alpha D t^{\alpha - 1}}{N^2} \sum\limits_{m=0}^{\infty} \:
(- 1)^m (2 m + 1) \: \exp\left[ - \dfrac{(2 m + 1)^2 \pi^2}{N^2} D \:
t^{\alpha} \right]
\label{FPT_Final}
\end{eqnarray}
The long time behavior of $Q$ is determined by the smallest eigenvalue in the
series eq.~(\ref{FPT_Final}). Therefore, at $t \to \infty$ one has
\begin{eqnarray}
  Q (t) = \dfrac{4\pi \alpha D t^{\alpha - 1}}{N^2} \exp\left( -
\dfrac{\pi^2}{N^2} D \: t^{\alpha} \right) .
\label{Long_Time}
\end{eqnarray}
i.e., the FPTD follows a stretched-exponential law at late times. We will try
to check this prediction in our MD-simulation study in Section \ref{sim}.

\subsection{Asymptotic behavior near the adsorbing boundary}

An interesting aspects of translocation dynamics in a system with two
adsorbing boundaries has been considered recently
\cite{Kantor_1,Kantor_2,Kantor_3}, suggesting that for a sufficiently long time
interval the normalized distribution
\begin{eqnarray}
p_{s_0}(s, t) =  \dfrac{W(s,s_0, t)}{\int_{0}^{N} W(s, s_0, t) d s},
\label{Stable_Shape}
\end{eqnarray}
reaches a stable (time-independent) shape which differs from the simple
$\sin$-function and at $s \to 0$ (i.e., close to the adsorbing boundary) the
function  $p_{s_0}(s) \sim s^{\phi}$ where $\phi > 1$ (for a $\sin$-function it
would have been $\phi = 1$).

Based on numerical results \cite{Kantor_1,Kantor_2}, one has tried \cite{Zoia}
recently to link translocation dynamics to self-affine processes undergoing
anomalous diffusion in bounded domains within the context of fBm. The
argumentation \cite{Zoia} relies on two crucial points: (i) The PDF $W(s, s_0,
t)$ has a self-affine form $W(s, s_0, t) = (1/t^{\alpha/2}) F(s/t^{\alpha/2},
s_0/t^{\alpha/2})$, and (ii) the so called {\it survival probability} $S (s_0,
t) = \int_{0}^{N} W(s, s_0, t) d s$  has a long-time scaling behavior $S (s_0,
t) \sim t^{- \theta}$ with an exponent $\theta = 1 - \alpha/2$. Simple scaling
arguments lead Zoia {\emph et. al.} \cite{Zoia} to the conclusion that the normalized probability
$p_{s_0}(s, t) = (1/t^{\alpha/2}) {\tilde p}_{y_0} (y)$, where $y =
s/t^{\alpha/2}$, $y_0 = s_0/t^{\alpha/2}$ and the scaling function ${\tilde
p}_{y_0} (y) \sim y^{\phi}$, with the  exponent $\phi = 2/\alpha - 1 \geq 1$.

Note that our solution, eq. (\ref{Final_Solution}), disagrees with the scaling
proposed in \cite{Zoia}. Since the eigenvalue spectrum is discrete
(as it should be for a finite interval $0 \leq s \leq N$), only the smallest
eigenvalue dominates in the long time limit so that for the PDF $W(s, s_0, t)$
and the survival probability $S(s_0, t)$ we get
\begin{eqnarray}
 W(s, s_0, t) &\sim& \left( \dfrac{2}{N}\right)  \sin \left(
\dfrac{\pi s_0}{N}\right) \: \sin \left(\dfrac{\pi s}{N}\right) \: \exp \left[-
\left(\dfrac{\pi}{N}\right)^2 D t^{\alpha} \right]\nonumber\\
S(s_0, t)&\sim& \left( \dfrac{4}{\pi}\right)  \sin \left(\dfrac{\pi
s_0}{N}\right) \: \exp \left[-\left(\dfrac{\pi}{N}\right)^2 D t^{\alpha} \right]
\label{W_S}
\end{eqnarray}
Therefore,
\begin{eqnarray}
 p_{s_0} (s, t) = \dfrac{W(s,
s_0, t)}{S(s_0, t)} = \dfrac{\pi}{2 N} \: \sin \left(\dfrac{\pi s}{N} \right)
\end{eqnarray}
Evidently, the self-affine scaling \cite{Zoia}, conjectured for $W(s, s_0, t)$
and $S(s_0, t)$, does not hold and the ``stable shape'' at $t \to \infty$ is a
simple $\sin$-function. This behavior is in accordance with the definition of
fBm \cite{West_Book}. The stretched exponential behavior for the survival
probability, which corresponds to eq. (\ref{W_S}), has been recently discussed
in ref. \cite{Oshanin}.

\section{Simulation results}
\label{sim}
To study the translocation process numerically, we performed Brownian Dynamics
(BD) simulations for two different chain lengths :  $N=51$ and $N=101$. The
polymer was modeled using a coarse-grained description in which the adjacent
monomers are connected  by finitely extensible nonlinear elastic (FENE) springs,
corresponding to a pair potential
\begin{align}
U_{FENE}(r_{ij})&=-\frac{k r_{ij}^2}{2}\ln\left(1-\frac{r_{ij}^2}{R_0^2}\right),
\end{align}
where $r_{ij}$ is the bond length between two beads and $R_0=1.5$ denotes its
maximal extension. All beads experience excluded volume interactions which are
modeled by a Lennard-Jones potential $U_{LJ}$, defined by
\begin{align}
 U_{LJ}(r_{ij}) = 4 \epsilon \left [\left( \frac{\sigma}{r_{ij}} \right)^{12}
- \left(\frac{\sigma}{r_{ij}} \right)^{6} \right],
\end{align}
where we use a cut-off $r_c = 2^{-1/6}\sigma$, implying that $U_{LJ} = 0$ for
$r_{ij} >r_c$. The parameter values were taken as $\epsilon = 1.0, \sigma =
1.0$, $k=30.0$, and were kept fixed during the simulations. The friction
parameter $\xi$ was taken as $\xi = 100 \sqrt{m \epsilon}/\sigma$, and the temperature
$T=1.2 \epsilon / k_B$, this implies a monomer diffusion coefficient $D_0 = k_B
T / \xi = 1.2 \epsilon / \xi$.

In order to simulate a translocation event, we create a separating membrane,
which consists of monoatomic layer, with a hole (the center monomer removed).
The membrane is placed inside a box of size (54$\times$54$\times$54) with
periodic boundary conditions and lies in the $z = 0$-plane. All atoms in the
membrane are frozen and interact with the translocating  chain via the Lennard-Jones potential. Before the translocation has
started, we put the middle bead of the chain symmetrically inside the pore, and
equilibrate the configuration, while keeping the middle bead fixed. To verify
that the chain has completely relaxed to equilibrium, we wait until the average
radius of gyration no longer changes with time. We have checked {\em a
posteriori}   that the number of translocation events to the \emph{trans} and
\emph{cis} side are evenly distributed.

\begin{figure}[htb]
\centerline{\includegraphics[scale=0.5]{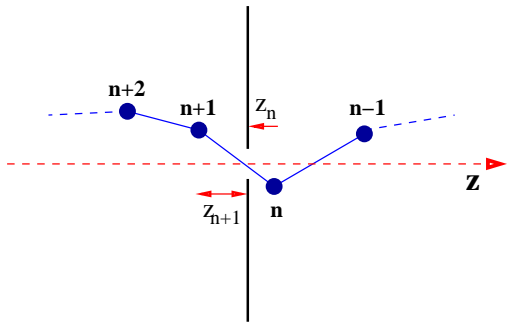}}
\caption{(Color online) The individual chain segments are enumerated starting from the
right-most terminal bead. The origin of the coordinate system is placed in the
pore. The $z$-coordinates of $n$ and $n+1$ beads are marked as $z_n$ and
$z_{n+1}$ respectively.} \label{Picture}
\end{figure}

\begin{figure}[htb]
 \includegraphics[width =8.0cm]{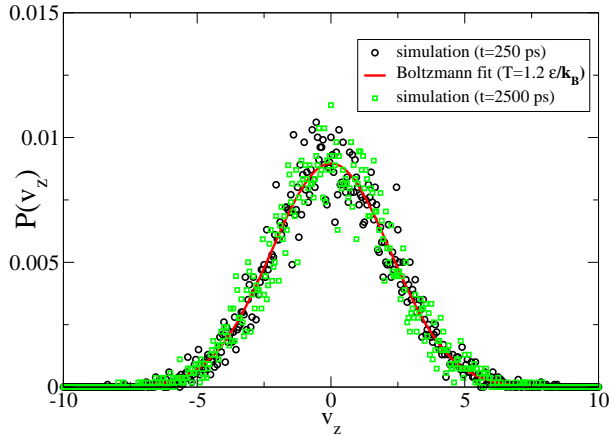}
\caption{(Color online) The velocity distribution function obtained from numerical simulations
at $T=1.2 \epsilon/k_B$, averaged over 9000 runs for two different times (t=250
ps and t=2500 ps). The simulations all collapse  on a single Maxwell-Boltzmann
distribution curve plotted in red.} \label{fig:veldisBD}
\end{figure}

\begin{figure}[ht]
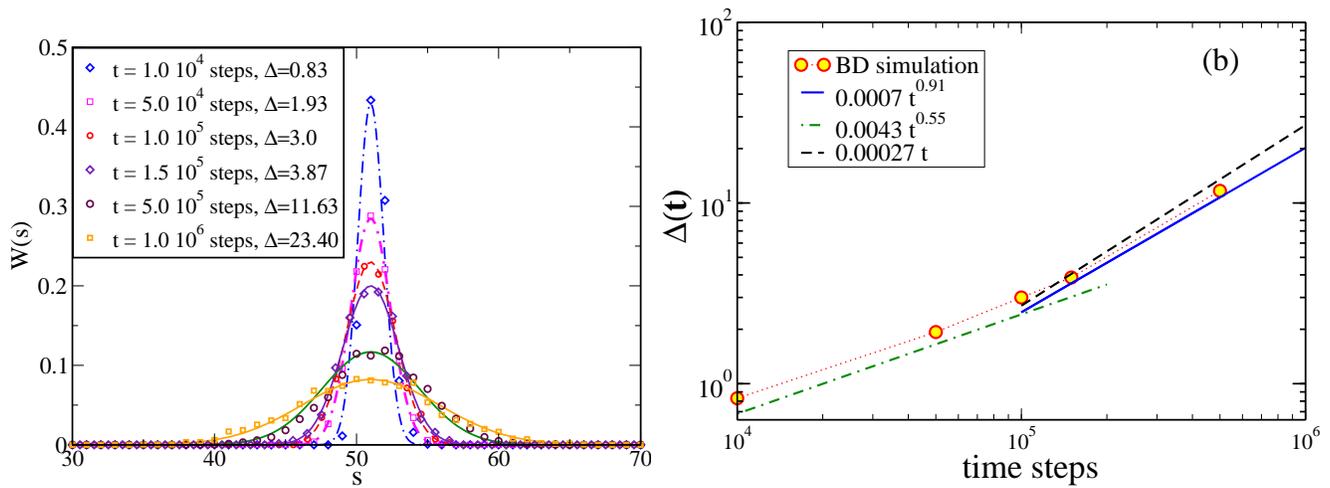

\vspace{1.0cm}
\includegraphics[scale=0.35, origin =br]{fig4a.eps}
\includegraphics[scale=0.35, origin=br]{fig4b.eps}
\caption{(Color online) (a) The  PDF  of the translocation coordinate $W(s, t)$ at 5 different
time moments. The 5 Gaussian fits correspond to the PDF of $s$  at  time steps $10^4$ till $10^6$ and have a variance 
$\Delta(t)$ whose value is given in the legend and which  increases with the number of time steps. (b) The
variance $\Delta(t)$ of the distribution $W(s,t)$  as a function of number of
time steps. It is clear that the behavior for short times is sub-diffusive and
scales as  $t^{0.55}$. For larger times  $t>10^5$ time steps, the behavior
changes and is only slightly sub-diffusive and  $\Delta(t)\simeq t^{0.91}$}
\label{fig:sdisBD}
\end{figure}

The translocation process of the polymer chain is studied by Brownian dynamics
simulations in a fictitious solvent. The translocation velocity $v(t)$ is sometimes 
approximated by $v_z(t)$, the $z-$component of the velocity of the bead inside the pore. From 
the recorded  data of the translocation coordinate and the velocity $v_z(t)$
we make histograms for the velocity- and the $s-$coordinate distributions; see also Figs.~\ref{fig:veldisBD}
and \ref{fig:sdisBD}a. It is
convenient to define the translocation coordinate $s(t)$ as a continuous
variable so that one may easily calculate the translocation velocity $v(t) = d
s(t)/d t$. In Fig.~\ref{Picture} we show a simple way to do this. The origin of
the coordinate system is located in the pore. The $z$-coordinates of the $n$-th
and $n+1$-th beads are marked as $z_{n}$ and $z_{n+1}$ respectively so that
$z_{n} \ge 0$ and $z_{n+1}\le 0$. The continuous translocation coordinate which
interpolates between two consecutive integer values $n$ and $n+1$ can be defined
as
\begin{equation}
s = \begin{cases} n+\frac{z_{n}}{|z_{n+1}|+z_n}& ,\text{if $z_n \ne 0$ and
$z_{n+1}\ne 0$,} \\
n& ,\text{if $z_n =  0$.}\\
n+1& ,\text{if $z_{n+1} =  0$.}
\end{cases}
\end{equation}

We plot the the velocity distribution function $P(v_z)$ (see
Fig.~\ref{fig:veldisBD} and the translocation coordinate distribution $W(s,
t)$ (Fig. \ref{fig:sdisBD} a) for several times $t=1.0\times10^4$,
$t=5.0\times10^4$, $t=1.0\times10^5$, $t=1.5\times 10^5$, and $t=5.0\times 10^5$
time steps. All distributions were obtained by averaging over at least $5000$
runs. Note that $1$ time step corresponds to $0.005$ ps = $0.005 \sigma \sqrt{m
/ \epsilon}$. The velocity distribution is clearly seen from
Fig.~\ref{fig:veldisBD} to be a Gaussian, centered around $v_z=0$. We also
plotted $P(v_z)$  for $t=50 000$ and $t=500 000$ time steps so as to demonstrate
that the distribution is indeed {\em time-independent}. Moreover, we verified
that the distribution is well described by a Maxwell-Boltzmann distribution for
the simulation temperature $T=1.2 \epsilon/k_B$. The PDF for the $s-$coordinate,
$W(s, t)$, is centered around $s=51$, which is the middle bead of the chain of
length $N=101$; the distribution broadens symmetrically.
From Figs. \ref{fig:veldisBD} and \ref{fig:sdisBD}a it can easily be seen   that
the distributions for both the velocity and the translocation coordinate $s$,
obtained from our Molecular Dynamics simulation are indeed Gaussian. In
Fig.~\ref{fig:sdisBD}b we show change of the variance $\Delta(t)$ (as defined by
eq. (\ref{Second_Moment})) of the Gaussian distribution curves $W(s,t)$  as a
function of time.

Evidently, $\Delta(t)$ behaves sub-diffusively as $\propto t^{0.55}$ for
shorter times, and later switches to a $t^{0.91}$-law for longer times. This
corresponds to the simple scaling consideration \cite{Chuang} which claims that
the variance of the translocation coordinate $\Delta(t) \equiv \langle
s^2\rangle - \langle s\rangle^2$ in the long time limit should go as $\Delta (t)
\propto t^{\alpha}$, where $\alpha = 2/(2 \nu + 1) \approx 0.92$ for $3D$-case.
This power is obtained by the requirement that the translocation time $\tau_{\rm
tr}$  should scale like the maximal relaxation time of the polymer chain, i.e.,
$\tau_{\rm tr} \propto N^{2\nu + 1}$. In addition, by the time when the
translocation is completed, the following relationship holds: $\Delta (\tau_{\rm
tr}) \propto N^2$.

We emphasize that the change in the exponent $\alpha$ from $0.55$ in the range
$[10^4,10^5]$ time steps to $0.91$ from $[10^5,10^6]$ time steps, is in
agreement with recently reported results by Amatai {\em et.
al.}~\cite{Kantor_3}, who considered the movement of a tagged monomer in a Rouse
chain. This somewhat surprising result indicates that the study of the  movement
of a tagged Rouse monomer can indeed provide essential  information about the
translocation process.

\begin{figure}[htb]
\vspace{1.0cm}
\centering
 \includegraphics[width = 0.45\textwidth]{fig5aREV.eps}
\vspace{1.0cm}
\includegraphics[totalheight = 0.450\textwidth, origin = br ,angle =270 ]{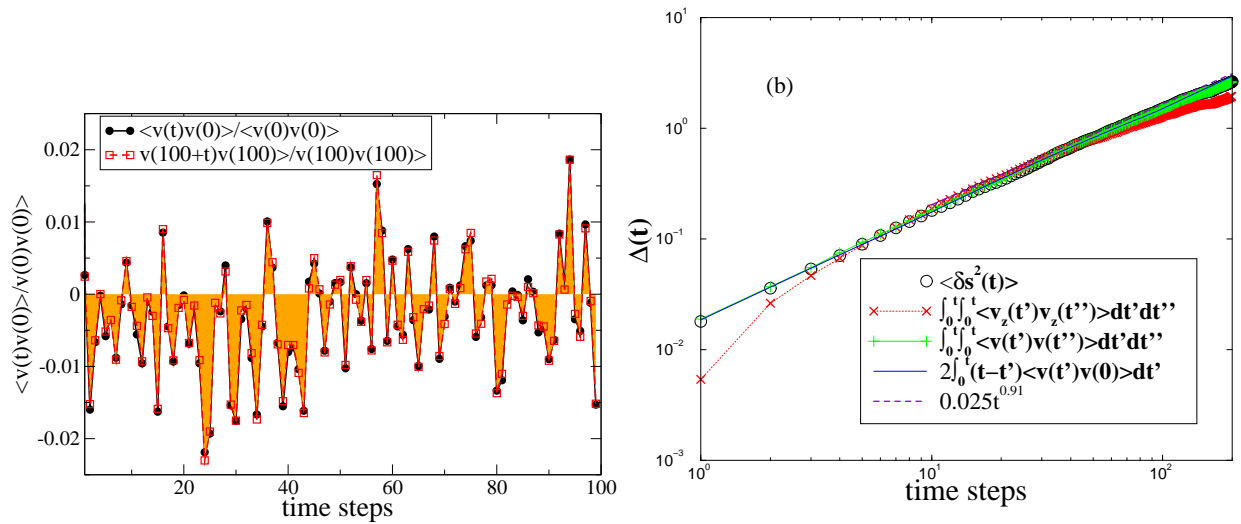}
\caption{(Color online) (a) The normalized velocity correlator against elapsed time $t$. (b)
Comparison of the $\Delta (t)$ data (circles) with the result of integration
of the velocity correlator according to  eq.(\ref{Velocity}) (blue solid line).
The double time integral $\int_{0}^{t} d t' \int_{0}^{t'} d t''\langle
v(t')v(t'')\rangle$,  which is valid even when time translational invariance
does not hold, is given by green $+$ symbols.}\label{fig:cor}
\vspace{0.9cm}
\end{figure}

Next we consider the velocity correlation functions.  The normalized
auto-correlation function is depicted in Fig.~\ref{fig:cor}a. From this figure
one can infer that the correlation time is not zero albeit rather small. One
clearly sees anti-correlation in the interval $t \in (0,5)$, after which the
correlations vanish on the average. Note that such unusual 'negative'
correlations at short time lags have been observed recently in experimental
studies of the sub-diffusive motion of bacterial chromosomal loci through a
viscoelastic cytoplasm \cite{Weber}.  To illustrate the invariance with respect
to shifts in time, a time shifted correlation function is also displayed. Since
the curves collapse onto each other, this demonstrates that the correlation
function is indeed time invariant as required.  In Fig.~\ref{fig:cor}b we give
the consistency check: the translocation coordinate mean-square displacement
$\Delta(t)$ as derived from the MD-simulation is compared to the time integral
given by eq.(\ref{Velocity}). One can see very good agreement between these two
ways of $\Delta (t)$ calculation, which also suggests that the velocity
correlator (cf. Fig.~\ref{fig:cor}a) has been found with reasonable accuracy.

We have also relaxed the time translational invariance assumption and
represented $\Delta (t)$ as double integral $\int_{0}^{t} d t' \int_{0}^{t'} d
t''\langle v(t')v(t'')\rangle$. The results of these calculations basically
superimpose onto each other which indicates again that the time translational
invariance holds indeed. From Fig.~\ref{fig:cor} it can also be seen that the
$z$-component of the monomer velocity inside the pore provides a reasonable
measure for the translocation velocity although there definitely are minor
deviations between this velocity and the translocation velocity. In particular,
the variance of the tagged particle  velocity has a somewhat smaller power law
exponent than the real translocation velocity. However, combining
Fig.~\ref{fig:cor}b with Fig.~\ref{fig:sdisBD}b shows that the exponent $\alpha$
changes with time from $0.91$ on the very short time scales $[1,200]$ to about
$0.55$ for times on the order of $10^4$. This is again in agreement with the
prediction of \cite{Kantor_3} for a tagged Rouse monomer.

\begin{figure}[htb]
\vspace{0.70cm}
 \includegraphics[scale=0.33]{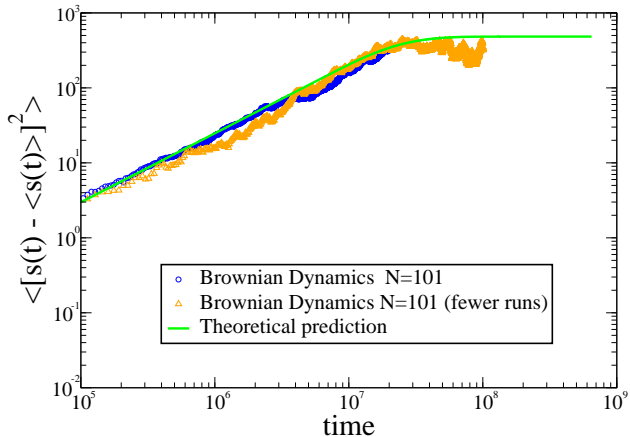}
\caption{(Color online) $\langle \Delta(t) \rangle$ for long times reaches a plateau. The
plateau height as well as the slope of the theoretical curve preceding the
plateau are in a good agreement with simulation results.}
\label{fig:second}
\vspace{1.0cm}
\end{figure}

Figure \ref{fig:second} demonstrates the $\Delta (t)$-behavior for a larger time
interval where it is seen to reach a plateau as soon as the polymer chain has
completed the translocation process. Here we compare the variance $\Delta (t)$
from numerical simulation with the analytical expression eq.
(\ref{Second_Moment}) where we took $\alpha = 0.92$. It can be clearly seen that
the results are in good agreement.

\begin{figure}[htb]
 \includegraphics[scale=0.35]{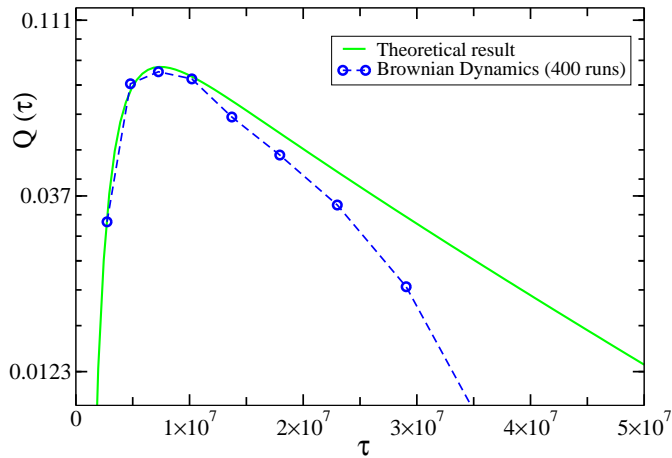}
\caption{(Color online) The FPT distribution $Q(\tau)$. Theory and simulations compared for a
N=51 chain.} \label{fig:tau}
\end{figure}

The first passage time distribution $Q(\tau)$ is interesting too. In a previous
paper \cite{Dubbeldam_1}, we employed the FFPE to describe the distribution
$Q(\tau)$. The tail of the distribution decreased as a power law in that case.
However, for the theory that we develop here, a stretched exponential decay is
expected. For a $N=51$ chain, we have performed simulations to compare $Q(\tau)$
with the theoretical prediction. For an average over $400$ runs, we obtained the
result  which is shown in Fig.~\ref{fig:tau}. The theoretical distribution is
found to be in reasonable agreement with the simulation result. We remark that
experimental and numerical  verification of the tail of the translocation time
distribution is extremely difficult, as the events constituting the
distribution's tail are very sparse.

\section{Conclusion}
\label{summary}

In the present work we treat the translocation dynamics within the framework of
{\em fractional} Brownian motion (fBm) whereby the definition of the
translocation coordinate velocity, $d s(t)/d t = v(t)$, is considered as a
Langevin equation where $v(t)$ can be viewed as a Gaussian process with given
two first cumulants: $\langle v(t)\rangle = f(t)/\xi_0$ and $ G (t) = \langle v
(t) v(0)\rangle - \langle v(t) \rangle^2 $. The first cumulant implies that
the friction $\xi_0$ is constant, in agreement with data from computer
experiments. This does not rule out the possibility that there are cases when
friction might be time-dependent \cite{Sakaue}. In such cases a modified version
of the governing equation (\ref{FP}) would be warranted.

Based on this Langevin equation, we derive (without any further assumption) the
appropriate Fokker-Planck equation of motion for the distribution function $W(s,
t)$ with a time-dependent drift and diffusion coefficient. The obtained solution
for $W(s,t)$ demonstrates two characteristic features of fBm which agree
favorably with the recent findings \cite{Kantor_1,Kantor_2,Kantor_3} as regards
nondriven translocation dynamics:
\begin{itemize}
 \item a Gaussian distribution for the translocation coordinate $s$ during the
translocation process
 \item a sub-diffusive behavior for the variance of the translocation
coordinate, $\Delta (t) = \langle  s^2 (t)\rangle - \langle s(t) \rangle^2$.
\end{itemize}
Moreover, the survival probability $S(t, s_0)$ in the long time limit has a
stretched-exponential form in agreement with recent findings \cite{Oshanin}
(albeit in contrast to a popular opinion about its scaling behavior
\cite{Zoia}). One should note, however, that the power ($0.92$) of the
stretched exponential function is very close to unity so that a distinction
from simple exponential decay of the survival probability at the long time
limit would be hardly detectable.

The velocity correlator $G(t)$ for the case of unbiased translocation has been
computed by means of Brownian dynamics simulationi and found to expose unusual
{\em negative} correlations in time. A consistency check, implemented by means
of eq.(\ref{Velocity}), shows that the calculated $G(t)$ is meaningful indeed.
Thus, it appears that the presence of negative velocity correlations reveals a
feature which is specific for the process of translocation in particular and
for anomalous diffusion in general \cite{Weber}.

We have calculated the variance of the translocation coordinate $\Delta (t)$
and shown that  $\Delta (t)$ follows a power law, with different exponents
$\alpha$, depending on the timescales one is interested in. In particular, for
times on the order of the translocation time, $\Delta \sim t^{0.91}$ in $3D$.
Furthermore, we examined the translocation time distribution function $Q(\tau)$,
derived from simulation results in a broad time interval and compared these to our
theoretical predictions. Unfortunately, attaining good statistics for very late
times still remains a difficult task for present days simulation studies so
more work will be needed in order to test the agreement with theoretical
predictions unambiguously.

Eventually, we have shown in Section \ref{chain_dynamics} that our approach can
be used for the biased translocation as well. This investigation will be
reported on in a separate presentation.

\section*{Acknowledgment}

We dedicate this work to our late colleague and collaborator S. Kotsev who took
part in the beginning of our investigations. The authors are indebted to A.Y.
Grosberg for stimulating discussions. We gratefully acknowledge the SFB-DFG 625
project for financial support. AM appreciates hospitality during his stay at the
Max-Planck Institute for Polymer Research in Mainz. JLAD is grateful for the technical support 
of P. Theodorakis with the GROMACS computations.

\end{document}